\documentclass[aps,pre,twocolumn,superscriptaddress,preprintnumbers,showpacs,nofootinbib]{revtex4-1}

\usepackage{amsmath,amssymb}
\usepackage[dvipdf,dvips,dvipdfmx]{graphicx}
\usepackage{here}
\usepackage{color}

\newcommand{\cc}{\mathrm{c}}
\newcommand{\itr}{\mathrm{itr}}

\begin{document}

\title{Tensor Renormalization Group Algorithms with a Projective Truncation Method}

\author{Yoshifumi Nakamura}
\email[]{nakamura@riken.jp}
\affiliation{RIKEN Center for Computational Science,
 Kobe, 650-0047, Japan}

\author{Hideaki Oba}
\email[]{h\_oba@hep.s.kanazawa-u.ac.jp}

\author{Shinji Takeda}
\email[]{takeda@hep.s.kanazawa-u.ac.jp}
\affiliation{Institute for Theoretical Physics, Kanazawa University,
 Kanazawa 920-1192, Japan}

\date{\today}

\begin{abstract}
  We apply the projective truncation technique to the tensor renormalization group (TRG) algorithm
  in order to reduce the computational cost from $O(\chi^6)$ to $O(\chi^5)$, where $\chi$ is the bond dimension,
  and propose three kinds of algorithms for demonstration.
  On the other hand, the technique causes a systematic error due to the
  incompleteness of a projector composed of isometries,
  and in addition requires iteration steps to determine the isometries.
  Nevertheless, we find that the accuracy of the free energy for the Ising model on a square lattice
  is recovered to the level of TRG with a few iteration steps even
  at the critical temperature for $\chi=32$, $48$, and $64$.
\end{abstract}

\preprint{KANAZAWA-18-04}


\maketitle

\section{Introduction}
Tensor networks are known to be a very useful method to study
the quantum and classical many-body systems.
For the quantum many-body system, the tensor network is
used to express the low-energy states as a trial wave function
for a variational method \cite{DMRG,DMRGreview,MPS,PEPS}.
On the other hand, the partition function of the classical lattice model and
Euclidean path integral of the quantum system on a lattice
are represented by the tensor network \cite{TNrepresentation,Liu:2013nsa}.
In such cases, the evaluation of the partition function or Euclidean path integral
is, in general, very hard.
The tensor renormalization group (TRG) method \cite{trg}, however, was invented
in order to efficiently carry out the computation, and
it is regarded as one of the real-space renormalization group methods. 
The methodology has been improved in a various ways on the basis of various philosophies \cite{SRG,TEFR,TNR,loopTNR,rec_trg_1,rec_trg_2,rec_trg_4}.

A striking feature of the tensor network method is that it is free of the sign problem.
With this property, the method attracts attention beyond condensed-matter physics
\cite{
Banuls:2013jaa,Kuhn:2015zqa,Banuls:2015sta,Banuls:2016lkq,Banuls:2016gid,Banuls:2017ena,
Shimizu:2012zza,
Shimizu:2014uva,
Shimizu:2014fsa,
Shimizu:2017onf,
Denbleyker:2013bea,Yu:2013sbi,Meurice:2016mkb,
Takeda:2014vwa,Kawauchi:2016xng,Sakai:2017jwp,Kadoh:2018hqq}.
There are, in fact, many interesting models which are suffering from the sign problem in high-energy physics:
QCD with finite quark density, the $\theta$ vacuum of QCD, chiral gauge theory, the super-symmetric model, and so on.
In order to study such models, a tensor network scheme for higher-dimensional systems is indispensable,
and it is known as the higher-order tensor renormalization group \cite{hotrg,hotrg2}.
For such higher-dimensional systems, however,
the computational complexity gets worse;
the cost is proportional to $\chi^{4d-1}$, where $d$ is the dimensionality of
a system and $\chi$ is the bond dimension of a tensor.
Thus, it is vital to develop a technique to reduce the cost while keeping accuracy.
So far some approaches for the cost reduction have been proposed:
the Monte Carlo approach \cite{VariationalQMC}, which
randomly samples an index of the tensor in a contraction process,
and the projective truncation technique \cite{isometry}
developed in the course of the tensor network renormalization method \cite{TNR}.

In this paper, we focus on the projective truncation technique
which inserts a projector consisting of a pair of isometries into a target local network.
The isometry is optimally determined by minimizing the proper cost function.
We apply the technique to TRG
and propose algorithms which reduce the cost from $O(\chi^6)$ to $O(\chi^5)$.
Furthermore we numerically examine the accuracy of a physical quantity and
measure the elapsed time to see its performance.

The rest of the paper is organized as follows.
In Sec.~\ref{sec:algorithms}, after briefly reviewing the original TRG algorithm, we explain
the projective truncation technique
and discuss its relation to the randomized singular-value decomposition (RSVD).
Then we propose three algorithms whose cost is reduced compared with the original TRG.
In Sec.~\ref{sec:numerical_results},
we present numerical results of the free energy of the Ising model on the square lattice
for a comparison between our algorithms and TRG
and discuss their performance.
Concluding remarks are given in Sec.~\ref{sec:conclusion}.
In the Appendix \ref{sec:det_iso}, we provide some details on how to determine
the isometry required in the projective truncation technique.

\section{Algorithms \label{sec:algorithms}}
In this section, 
after briefly summarizing the original TRG algorithm,
we explain the projective truncation technique 
and discuss its relation to the first stage of the RSVD.
Then, we propose algorithms using the technique and show
that the cost of the original TRG $O(\chi^6)$ is reduced to $O(\chi^5)$.
Although there are a variety of ways to exploit the technique,
here we present three algorithms for demonstration.

\subsection{Tensor renormalization group\label{subsec:TRG}}

Let us begin with an explanation of the TRG algorithm \cite{trg,TERG} for the square lattice.
First of all, we rewrite the partition function in terms of a tensor network,
\begin{align}
  Z = \sum_{\ldots ,i,j,k,l,m,n,o, \ldots} \cdots T^{(\mathrm{init})}_{ijkl}T^{(\mathrm{init})}_{mnio} \cdots,
\end{align}
where $T^{(\mathrm{init})}$ is an initial tensor\footnote{
The initial tensor is model dependent, and an actual procedure
to create it also depends on the details of a model,
physical degrees of freedom (scalar fields, fermion fields, or
gauge fields), and forms of interaction (hopping term, plaquette loop).
We, however, will not go into this but stick with the Ising model in the following.
},
which is an ingredient of the tensor network, and has
indices $i,j,k,l$ running from 1 to $D^{(\mathrm{init})}$, which is the bond dimension of the initial tensor.
The tensor and its indices are located on a lattice point
and bonds of the lattice, respectively.

\begin{figure}[t]
\centering
 \includegraphics[width=60mm]{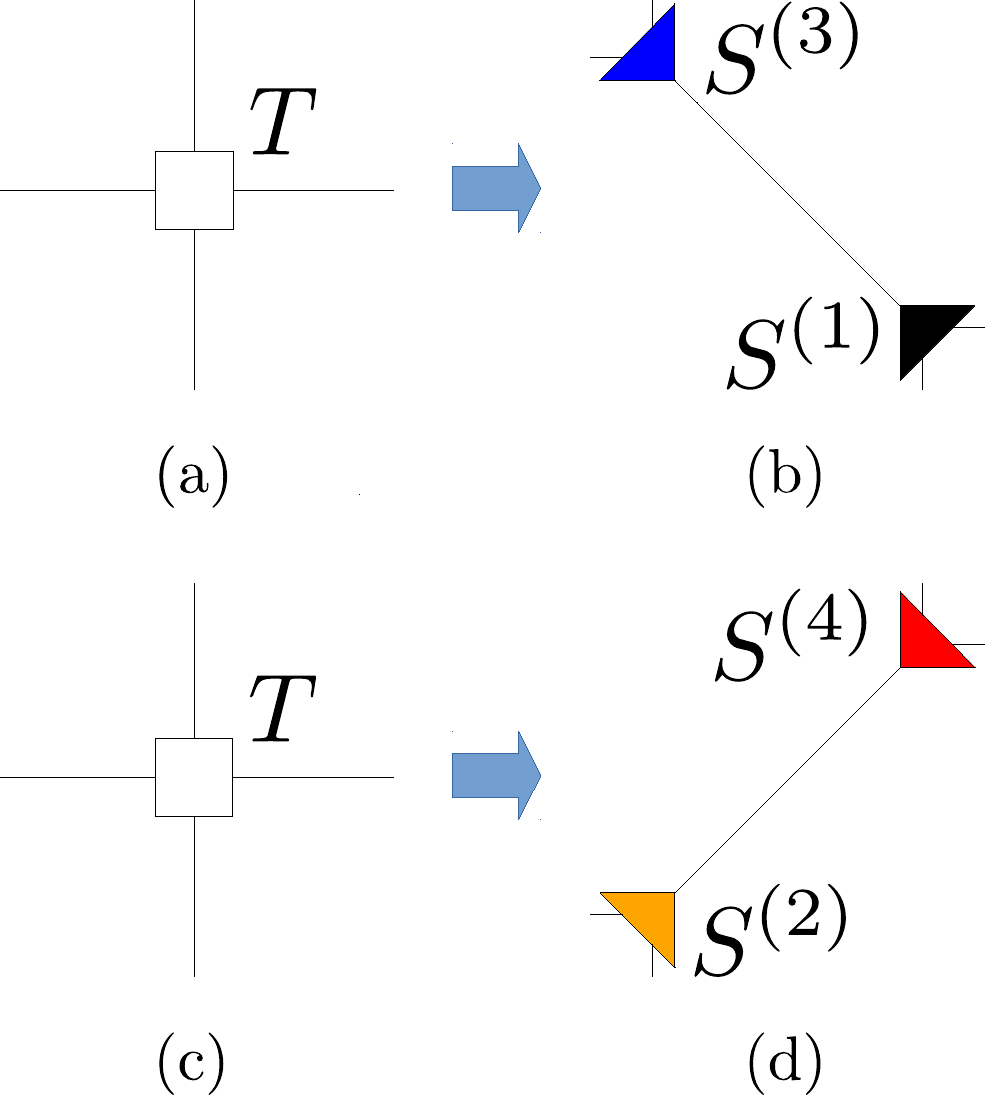}
 \caption{TRG decomposition.
   (a) The white square represents a four-leg tensor $T$.
   From (a) to (b) the four-leg tensor is decomposed into two three-leg tensors in Eq. (\ref{eq:S13}).
   (b) The triangles represent three-leg tensors, $S^{(1)}$ and $S^{(3)}$.
   Going from (c) to (d) shows decomposition of $T$ to $S^{(2)}$ and $S^{(4)}$ in Eq. (\ref{eq:S24}).
 }
 \label{fig:trg_dec}
\end{figure}

The next step is a coarse graining of the tensor network
to reduce the degrees of freedom.
This step can be divided into two parts: the first one is a decomposition
of the fourth-order tensor (TRG decomposition) and the second is a contraction to make 
a coarse-grained fourth-order tensor (TRG contraction).
In the TRG decomposition part,
the SVD is usually used,
and there are two ways to decompose the fourth-order tensor into
two third-order tensors:
\begin{align}
  T_{ruld} &\simeq \sum_{m=1}^{\chi} S^{(1)}_{rdm} S^{(3)}_{lum},\label{eq:S13}\\
  T_{ruld} &\simeq \sum_{n=1}^{\chi} S^{(2)}_{ldn} S^{(4)}_{run}\label{eq:S24},
\end{align}
where the original indices $r,u,l,d$ run from 1 to $D$ (see Fig.~\ref{fig:trg_dec}).
The overall range of the new indices $m,n$ is from 1 to $D^2$, but
the sum retains only the $\chi(\le D^2)$ largest singular values,
and this dictates the degree of the low-rank approximation of SVD.

\begin{figure}[t]
\centering
 \includegraphics[width=60mm]{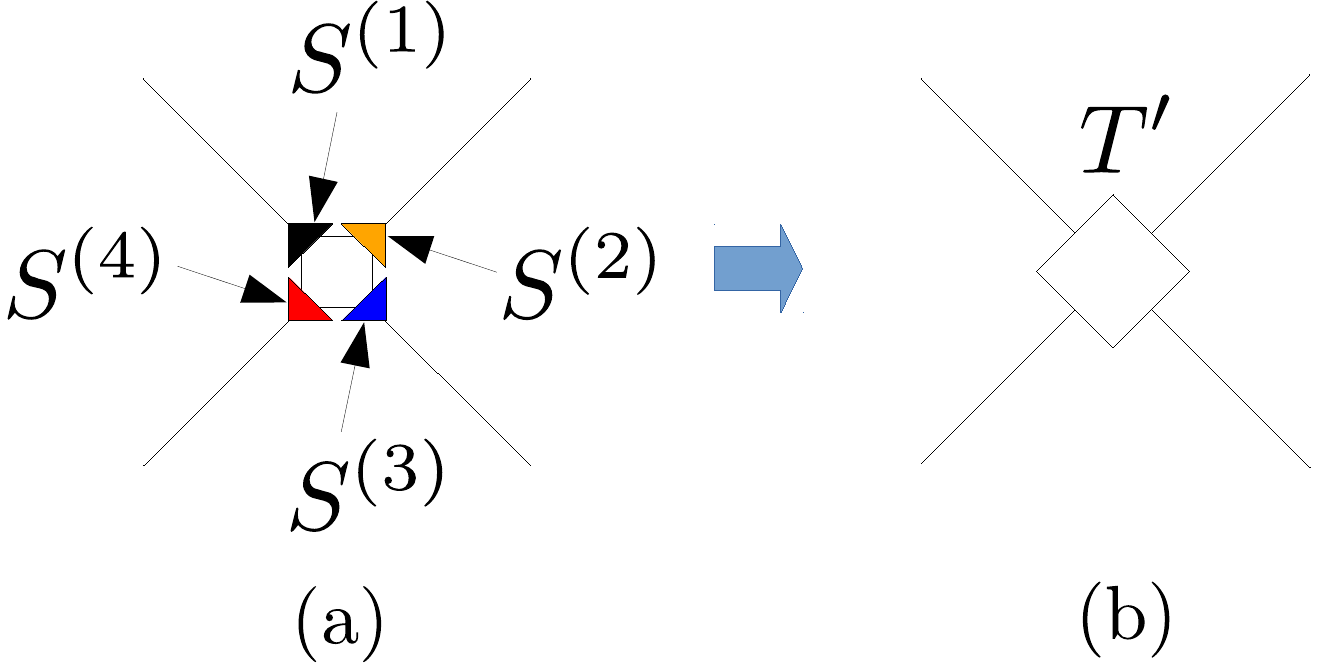}
 \caption{TRG contraction.
 From (a) to (b) contraction of four three-leg tensors gives a new four-leg tensor $T^\prime$. 
 }
 \label{fig:trg_cont}
\end{figure}

\begin{figure}[t]
\centering
 \includegraphics[width=80mm]{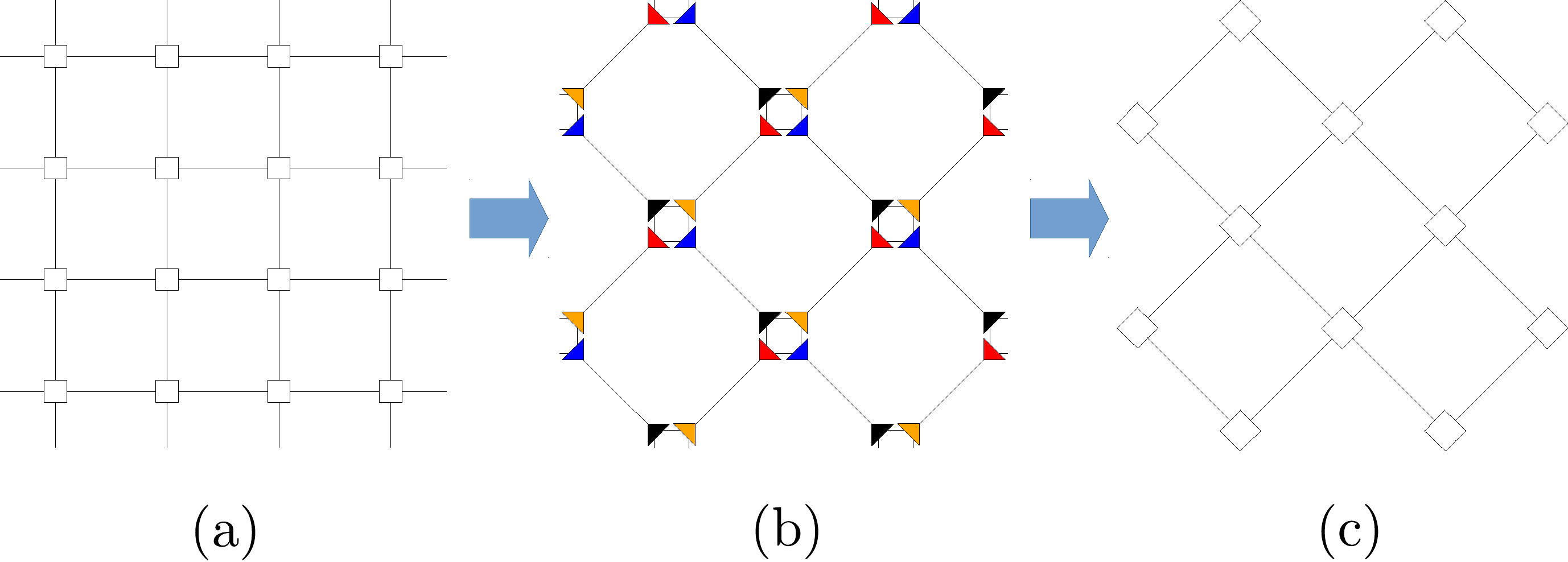}
 \caption{The coarse graining of the tensor network by TRG.
 From (a) to (b), TRG decomposition.
 From (b) to (c), TRG contraction.}
 \label{fig:trg}
\end{figure}

The second part is the TRG contraction: a contraction of four third-order tensors to make a new tensor
(see Fig.~\ref{fig:trg_cont})
\begin{align}
  T'_{ruld} = \sum_{\alpha, \beta, \gamma, \omega =1}^D 
  S^{(1)}_{\omega \alpha u} S^{(2)}_{\omega \beta r} S^{(3)}_{\gamma \beta d} S^{(4)}_{\gamma \alpha l}.
\label{eqn:contraction}
\end{align}
This calculation is done exactly.
The coarse graining of the network by TRG is shown in Fig.~\ref{fig:trg}.
By repeating the coarse graining, the number of tensors is reduced,
and one can obtain a network with a sufficiently small number of tensors.
The final step is to evaluate the partition function by
contracting such a network including a few tensors.

Here we remark on the cost of TRG.
In the TRG decomposition, the cost of SVD is $O(\chi^6)$.
This, however, can be reduced to $O(\chi^5)$ if one uses
the partial SVD (truncated SVD) \cite{PSVD} or the randomized SVD \cite{RSVD,RSVDTRG}.
Thus, this is not a crucial part.
A non trivial one is the TRG contraction in Eq. (\ref{eqn:contraction}), whose cost is $O(\chi^6)$.
As long as one adheres to the exact calculation, it seems hard to reduce the cost.
In order to break this situation,
we shall attempt to reduce the cost by applying the projective truncation technique
which approximately calculates a tensor contraction. 
The details of the method are explained in the following.

\subsection{Projective truncation technique and its relation to randomized SVD \label{subset:PT}}
In this section, we explain the projective truncation technique and
discuss its relation to the first stage of the randomized SVD \cite{RSVD}.

\begin{figure}[t]
\centering
 \includegraphics[width=50mm]{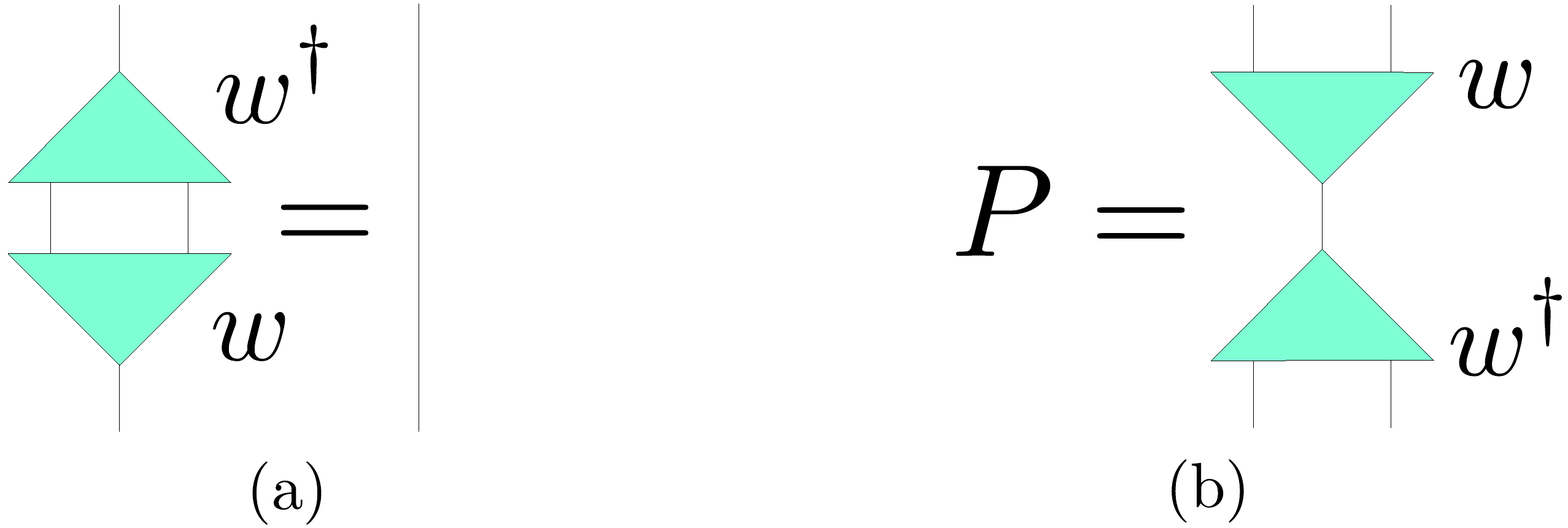}
 \caption{The properties of isometry. 
   (a) The two-leg contraction between $w$ and its conjugate gives an identity $w^\dag w = \mathbb{I}$. 
   (b) The one-leg contraction between them turns out to be a projector $w w^\dag = P$,
   and it satisfies $P=P^2$ because of the property in (a). }
 \label{fig:property_of_isometry}
\end{figure}

First, we introduce an isometry $w$, a $\chi\times \chi\times \chi$ tensor ($\chi^2\times \chi$ matrix) which satisfies $w^\dag w = \mathbb{I}$, where
$\mathbb{I}$ is a $\chi\times \chi$ identity matrix.
Figure \ref{fig:property_of_isometry} shows the properties of the isometry.
In the rest of the paper,
we assume that a single line for a bond in a figure represents an index running from 1 to $\chi$.
An isometry has often been used in coarse-graining procedures, but here
it is used for the purpose of reducing the computational cost of contraction for a given local network.
$P=w w^\dag$ is a $\chi\times \chi\times \chi\times \chi$ tensor ($\chi^2\times \chi^2$ matrix)
and has a property of the projector ($P=P^2$).
The concept of the projective truncation technique is well organized in \cite{isometry},
and its basic idea is that an original local network ${\cal N}$
is approximately replaced by a new one, $\tilde {\cal N}={\cal N}P$,
containing a projector $P$.
An isometry in the projector is determined by minimizing a cost function,
\begin{align}
\delta 
= \frac{||{\cal N}-{\cal N}P ||}{||{\cal N}||}
= \frac{||{\cal N}-{\cal N}ww^\dag ||}{||{\cal N}||},
\label{eq:delta_general}
\end{align}
where $||\cdots||$ is the Frobenius norm of a tensor.
One can iteratively solve the problem
after a linearization of the cost function (namely, fixing its conjugate $w^\dag$) \cite{VidalEvenbly,isometry},
although this is originally a non linear problem with regard to $w$.
Thus, when using the projective truncation technique,
one has to treat an additional parameter, an iteration number $n_\itr$.
For minimization of Eq. (\ref{eq:delta_general}),
the best choice of $w$ is the right singular vectors of ${\cal N}$, which correspond to the leading $\chi$ singular values. 
Therefore, $\delta$ may not become zero even if the iteration number is sufficiently large
\footnote{
The size of error by the local approximation
depends on the original local network itself.}.
Some details of the actual procedure to determine
the isometry are summarized in the Appendix \ref{sec:det_iso}.

Here we comment on the relation between the first stage of the RSVD
and the projective truncation technique.
First, let us recall the RSVD.
It can efficiently carry out the SVD for an $m\times n$ matrix $A$ only with
leading $k$ singular values and corresponding singular vectors.
The first stage of the RSVD is to find a basis matrix $Q$, which is an $m\times(k+p)$ matrix,
and column vectors are orthogonal $Q^\dag Q=1$ for the target matrix $A$
by minimizing the cost function
\begin{align}
||A-QQ^\dag A||.
\label{eq:RSVD1}
\end{align}
Here the oversampling parameter $p$ is introduced, and this dictates the accuracy of the RSVD.
To obtain the basis matrix $Q$, first of all, one prepares an $n\times(k+p)$ random matrix $\Omega$ as an input.
Then after forming $Y=A\Omega$, $Q$ is obtained by QR decomposition of $Y=QR$.
One notices a similarity between Eq. (\ref{eq:delta_general}) for a tensor network ${\cal N}$ and Eq.(\ref{eq:RSVD1}) for a matrix $A$ ,
and the basis matrix $Q$ corresponds to the isometry $w$.
On the other hand, the way to determine $Q$ is
different from that of the isometry $w$,
and $Q$ has an oversampling parameter $p$, while
$w$ is considered a fixed dimensionality, say a $\chi^2\times\chi$ matrix.
Although it is interesting to compare the performance of the two methods systematically,
we leave it for future works.

\subsection{Projectively truncated TRG \label{subsec:PTTRG}}

\begin{figure}[t]
\centering
 \includegraphics[width=80mm]{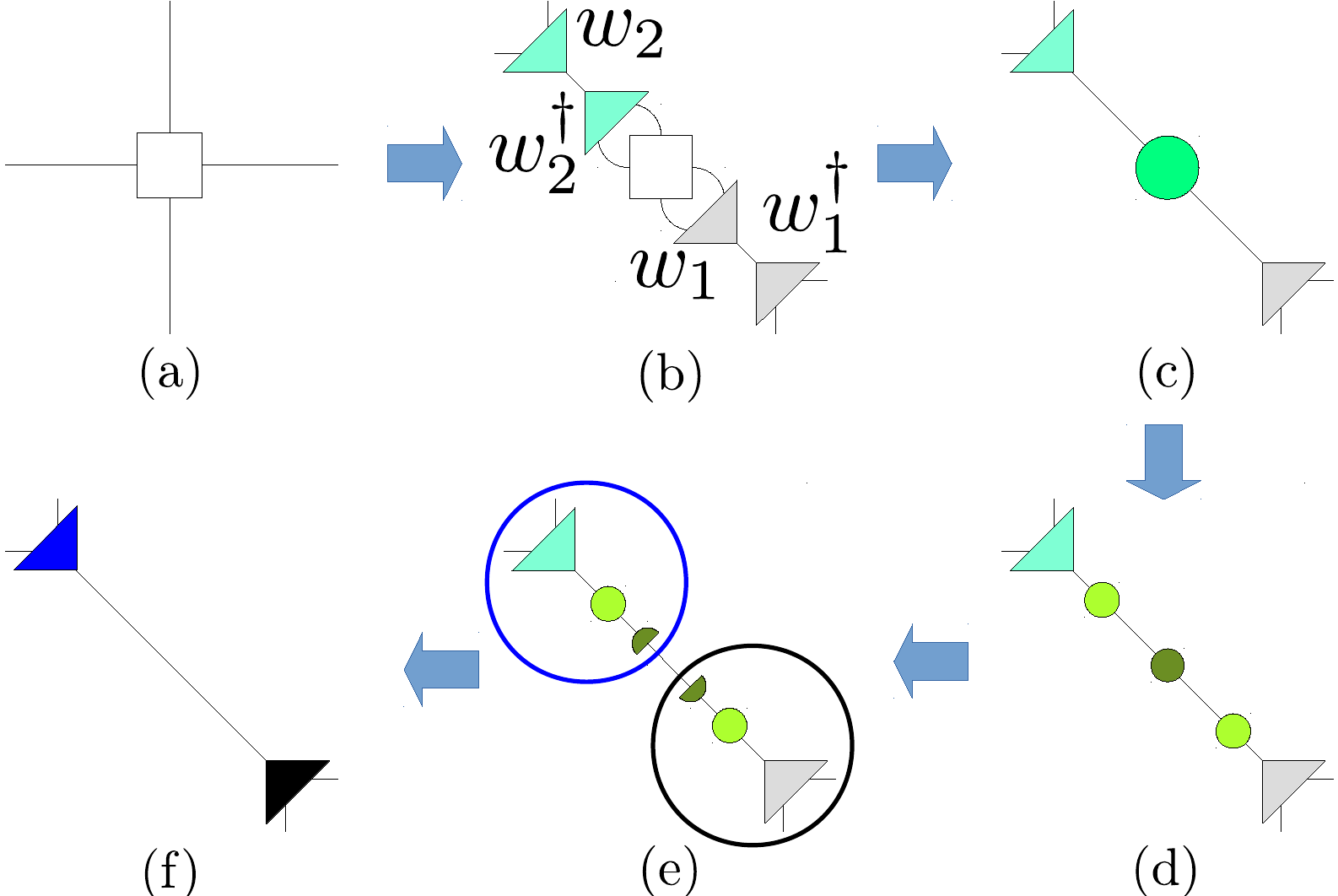}
 \caption{
 PTTRG decomposition: Decomposing the four-leg tensor into two three-leg tensors with the projective truncation technique.
 From (a) to (b) two projectors, $w_1w_1^\dag$ and $w_2w_2^\dag$, are inserted.
 From (b) to (c) contraction of $w_1$, $w_2^\dag$, and $T$ gives the $\chi \times \chi$ matrix (green circle).
 From (c) to (d) SVD of the matrix.
 From (d) to (e) dividing the singular values.
 From (e) to (f) contraction within a circle yields a three-leg tensor. Finally, $S^{(1)}$ and $S^{(3)}$ are obtained.
  }
 \label{fig:pttrg_dec}
\end{figure}

Now let us apply the projective truncation technique to TRG.
Figure \ref{fig:pttrg_dec} shows a decomposition of $T$ into $S^{(1)}$ and $S^{(3)}$
with the use of the technique.
For an original local network (a single tensor $T$) in Fig.~\ref{fig:pttrg_dec}(a),
two projectors $w_1w_1^\dag$, and $w_2w_2^\dag$, are inserted as shown in Fig.~\ref{fig:pttrg_dec}(b).
The isometries $w_1$ and $w_2$ are determined by solving the problem
\begin{align}
  \min_{w_1,w_2} \delta_w &= \min_{w_1,w_2} \frac{||T - w_2w^\dag_2 T w_1w^\dag_1|| \label{eq:delta_w}}{||T||}.
\end{align}
One can treat it as a linear problem with respect to $w_1$ for fixed $w_1^\dag$, $w_2$, and $w_2^\dag$
(see the Appendix \ref{sec:det_iso}).
An update of $w_2$ can be done in the same way.
Contraction of $w_1$, $w_2^\dag$, and $T$ in Fig.~\ref{fig:pttrg_dec}(b) gives a network
in Fig.~\ref{fig:pttrg_dec}(c), where a green circle is a $\chi\times \chi$ matrix.
The $\chi\times\chi$ matrix (green circle) is decomposed by SVD, as shown in Fig.~\ref{fig:pttrg_dec}(d),
where the light-green circles represent unitary matrices and the dark-green circle
is a diagonal matrix whose diagonal elements are singular values.
In Fig.~\ref{fig:pttrg_dec}(e),
the diagonal matrix is decomposed by the square root, and the decomposed matrix is denoted by
a half circle. 
By integrating the subnetworks with the black circle and the blue one in Fig.~\ref{fig:pttrg_dec}(e),
one obtains $S^{(1)}$ and $S^{(3)}$, respectively, in Fig.~\ref{fig:pttrg_dec}(f).
Similarly, one can obtain $S^{(2)}$ and $S^{(4)}$.
In this way we obtain $S^{(1,2,3,4)}$ from a tensor $T$ by using the projective truncation technique.
We call this procedure a projectively truncated TRG (PTTRG) decomposition.
This procedure may be equivalent to the truncated SVD.
Note that the computational cost of the PTTRG decomposition (all contraction processes
in Fig.~\ref{fig:pttrg_dec} as well as the determination of the isometries $w_1$ and $w_2$)
is $O(\chi^5)$.
As mentioned before, in order to realize the cost reduction of th order of $\chi^5$ in the decomposition part,
one may use the partial SVD (truncated SVD) instead, but
in this paper we persist with the projective truncation technique shown here.

\begin{figure}[t]
\centering
 \includegraphics[width=80mm]{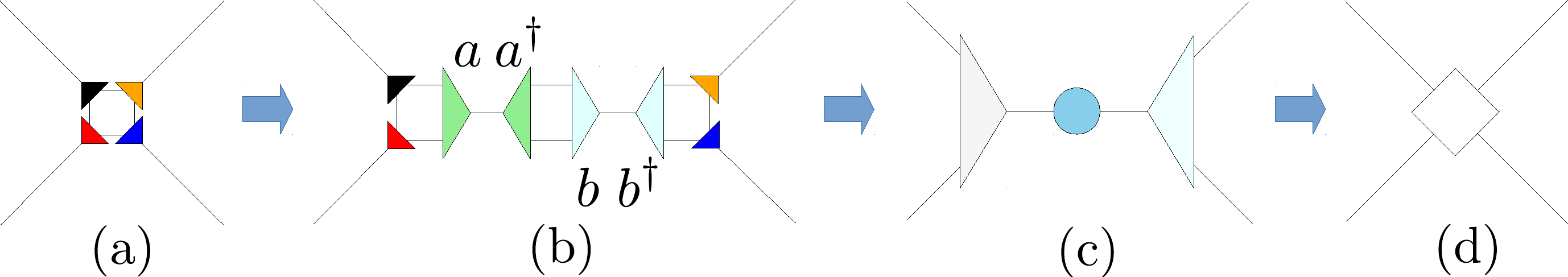}
 \caption{PTTRG contraction: Forming a four-leg tensor from four three-leg tensors by inserting projectors.
 From (a) to (b) two projectors, $aa^\dag$ and $bb^\dag$, are inserted.
 From (b) to (c) $S^{(1)}$, $S^{(4)}$, and $a$ are contracted; then a three-leg tensor is obtained.
 $S^{(2)}$, $S^{(3)}$, and $b^\dag$ are contracted to make a three-leg tensor.
 $a^\dag$ and $b$ are contracted; then a matrix (blue circle) is obtained.
 From (c) to (d) contraction of two three-leg tensors and the matrix gives a coarse-grained tensor $T'$.}
 \label{fig:pttrg_cont}
\end{figure}

Next, let us see how to use the projective truncation technique in the TRG contraction part
(PTTRG contraction).
Figure \ref{fig:pttrg_cont} shows the flow of this part.
From Fig.~\ref{fig:pttrg_cont}(a) to Fig.~\ref{fig:pttrg_cont}(b), two projectors, $aa^\dag$ and $bb^\dag$,
 are inserted
along the horizontal direction, and one may divide the network into two subnetworks:
one is $aa^\dag$, $S^{(1)}$, and $S^{(4)}$, and
the other is $bb^\dag$, $S^{(2)}$, and $S^{(3)}$.
Isometry within each subnetwork is independently determined by minimizing the following cost functions:
\begin{align}
  \delta_a &= \frac{||S^{(1)}S^{(4)} - S^{(1)} S^{(4)}aa^\dag||}{||S^{(1)}S^{(4)}||},\label{eq:delta_a}\\
  \delta_b &= \frac{||S^{(2)}S^{(3)} - S^{(2)} S^{(3)}bb^\dag||}{||S^{(2)}S^{(3)}||}.\label{eq:delta_b}
\end{align}
From Fig.~\ref{fig:pttrg_cont}(b) to Fig.~\ref{fig:pttrg_cont}(d), subnetworks are sequentially contracted
to make a coarse-grained four-leg tensor in the end.
Note that the computational cost in the process is kept to $O(\chi^5)$ at most.
One may insert the projectors along the vertical direction
and the procedure is basically the same as that of the horizontal case
shown in Fig.~\ref{fig:pttrg_cont}.

\begin{figure}[H]
\centering
 \includegraphics[width=80mm]{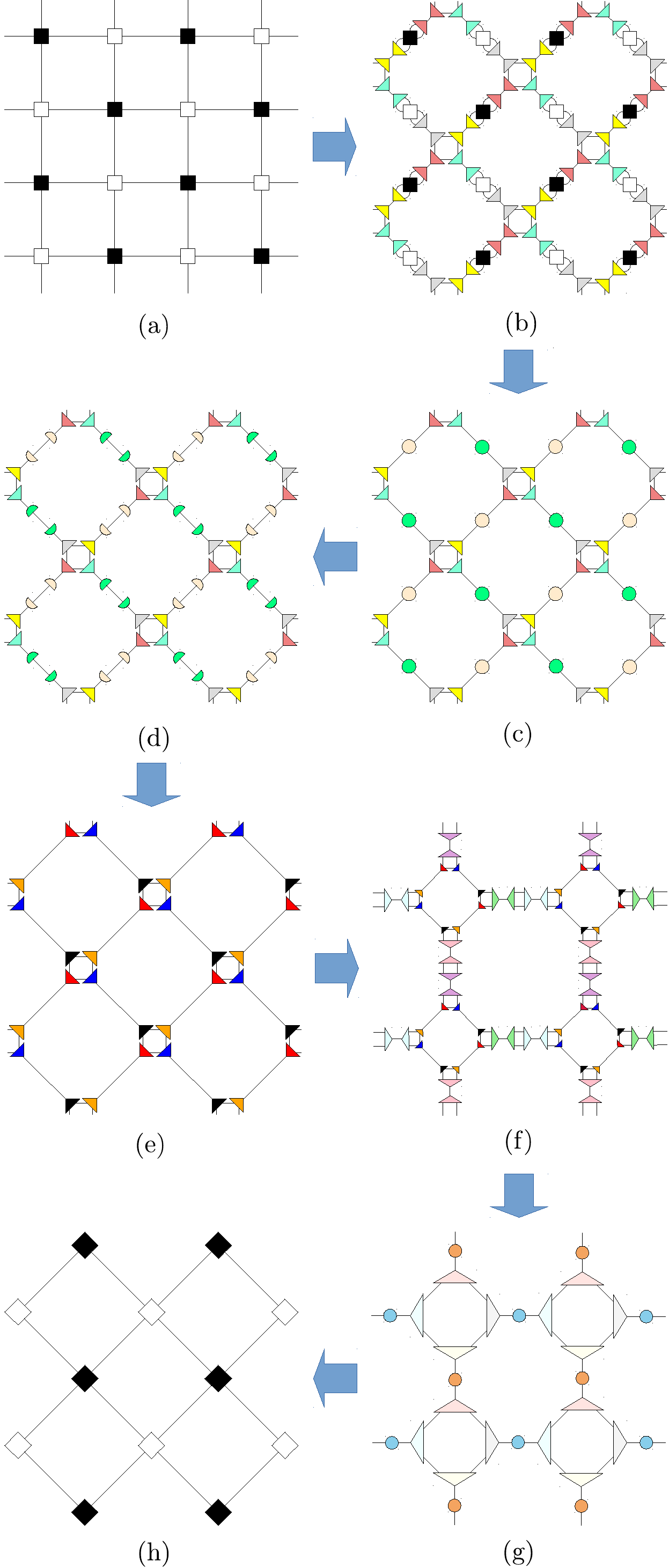}
 \caption{The whole procedure of PTTRG.
   (a) Tensor network consisting of two kinds of four-leg tensors (checkerboard structure).
   From (a) to (b) insertion of four kinds of projectors.
   From (b) to (c) contracting a four-leg tensor and two isometries gives a matrix (circle).
   There are two kinds of matrices.
   From (c) to (d) decomposing the matrices. Decomposed matrices are denoted by a half circle.
   From (d) to (e) contracting the matrix and the isometry gives a network similar
   to that in Fig.~\ref{fig:trg}(b). Completion of the PTTRG decomposition part.
   From (e) to (f) insertion of two kinds of projectors for each direction.
   In total, there are four kinds of projectors for the contraction part.
   From (f) to (g) contractions are done in subnetworks.
   The procedure for the horizontal direction is the same as that in Figs.~\ref{fig:pttrg_cont}(b) and \ref{fig:pttrg_cont}(c).
   The vertical direction is done as well.
   From (g) to (h) forming a four-leg tensor. The resulting network also has the checkerboard structure.}
 \label{fig:pttrg}
\end{figure}

The whole procedure of PTTRG is shown in Fig.~\ref{fig:pttrg}.
The starting network [Fig.~\ref{fig:pttrg}(a)] is assumed to be a checkerboard structure\footnote{
A uniform network is an element of a set of checkerboard networks.
As we will see, even if one starts the procedure with a uniform network,
the resulting network turns out to be a checker-board.
}
which contains two kinds of four-leg tensors.
In Fig.~\ref{fig:pttrg}(b), four kinds of projectors are inserted (two projectors for each four-leg tensor).
Two isometries and one four-leg tensor in Fig.~\ref{fig:pttrg}(b) are contracted,
and a matrix (circle) is obtained in Fig.~\ref{fig:pttrg}(c).
There are two kinds of matrices.
The matrices are decomposed by SVD [Fig.~\ref{fig:pttrg}(d)] and
the decomposed matrices are represented by half circles.
Contraction between the matrix and the isometry gives Fig.~\ref{fig:pttrg}(e),
which has the same structure as that of the TRG decomposition in Fig.~\ref{fig:trg}(b).
In Fig.~\ref{fig:pttrg}(f), four kinds of projectors are inserted in both the horizontal and vertical directions.
By carrying out the contraction sequentially, via Fig.~\ref{fig:pttrg}(g),
one arrives at Fig.~\ref{fig:pttrg}(h), which has a checkerboard structure.
Returning to Fig.~\ref{fig:pttrg}(a) with a rotation, one can restart the procedure,
and then the coarse graining is repeated.
Note that the computational cost of PTTRG is of the order of $\chi^5$.
The determination of an isometry requires $n_{\rm itr}$ iteration steps, and
the values of $n_\itr$ for the PTTRG contraction and decomposition steps are taken to be the same.
Thus, the total cost of the PTTRG is $O(n_{\rm itr}\chi^5)$.

Here we make a remark about the PTTRG algorithm.
First, let us consider two processes:
one is the PTTRG contraction from Fig.~\ref{fig:pttrg}(g) to Fig.~\ref{fig:pttrg}(h),
and the other is the PTTRG decomposition
from Fig.~\ref{fig:pttrg} [which is the same as Fig.~\ref{fig:pttrg}(a)]
to Fig.~\ref{fig:pttrg}(e) in the next coarse-graining step.

\begin{figure}[t]
\centering
\includegraphics[width=80mm]{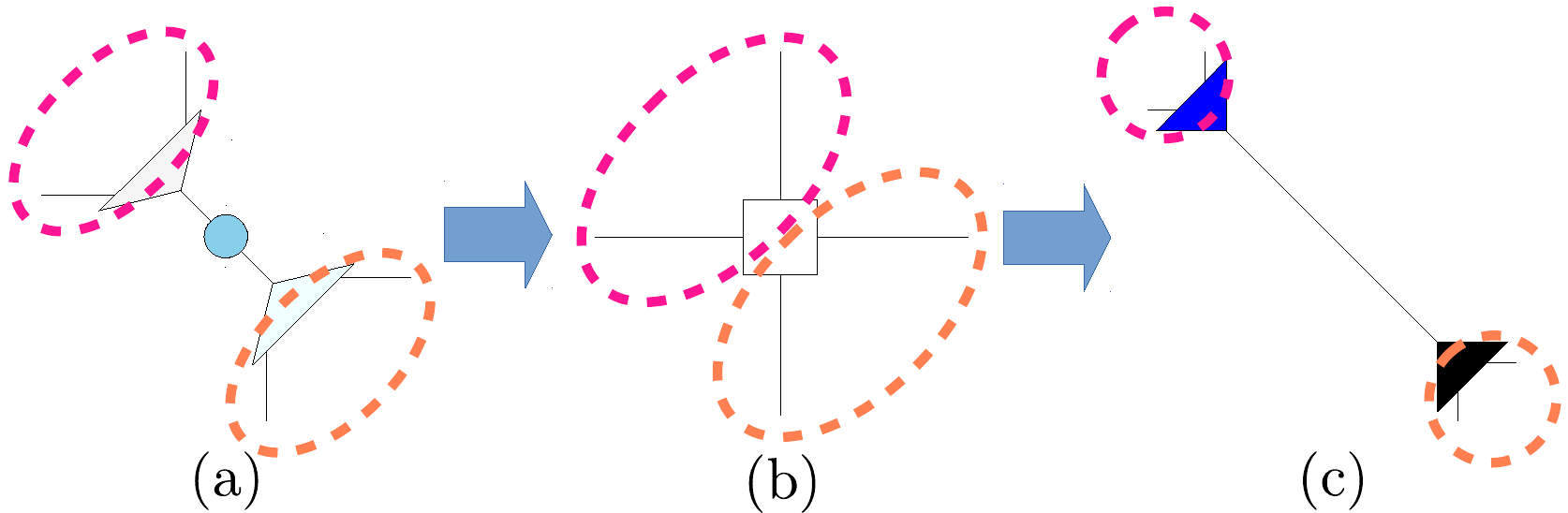}
 \caption{
 The direction of the contraction is the same as that of the decomposition; that is,
 the pair of non contracted indices of three-leg tensors in (a) and that in (c) are the same.
 From (a) to (b) PTTRG contraction. Contraction of the two three-leg tensors and the matrix gives
 the four-leg tensor whose rank is $\chi$.
 From (b) to (c) PTTRG decomposition in the next coarse-graining step. The four-leg tensor is decomposed into
 two three-leg tensors without loss of information.
 }
 \label{fig:pttrg_ll}
\end{figure}

In order to make the point clear,
we show local expressions for the two processes in Fig.~\ref{fig:pttrg_ll}.
An important point is that a pair of non contracted indices
(encircled bonds in Fig.~\ref{fig:pttrg_ll} form a pair of indices) is not
changed during the process from Fig.~\ref{fig:pttrg_ll}(a) to Fig.~\ref{fig:pttrg_ll}(c).
In other words, the direction of the contraction to make the new four-leg tensor and
that of the decomposition of the tensor are the same.
The other important point is that
the rank of a matrix associated with the four-leg tensor ($\chi^2\times\chi^2$ matrix) of PTTRG 
[Fig.~\ref{fig:pttrg_ll}(b)] is $\chi$
since it is basically made of two three-leg tensors [Fig.~\ref{fig:pttrg_ll}(a)]
that are a $\chi^2\times\chi$ matrix whose rank is $\chi$.
Therefore, the four-leg tensor is decomposed into the two new three-leg tensors without the loss of information
[Fig.~\ref{fig:pttrg_ll}(c)] as long as the direction of the decomposition is the same
as that of the contraction.
In this way the information reduction from $\chi^2$ to $\chi$ takes place in the
contraction part of PTTRG,
while for TRG it is done in the decomposition part where
the four-leg tensor, whose rank is $\chi^2$ in general,
is decomposed and only the rank $\chi$ parts are retained.
Although the timing of information reduction is different,
the network in terms of the three-leg tensors $S^{(1,2,3,4)}$ for PTTRG with sufficiently large $n_\itr$
has the same information as that of TRG.
Therefore, we conclude that
PTTRG converges to the normal TRG in the large-$n_\itr$ limit.

\subsection{PTTRG without forming four-leg tensors \label{subsec:PTTRG2}}
The PTTRG algorithm is motivated to reduce the cost of the TRG algorithm.
Since we give first priority to the understandability of the algorithm,
the algorithm shown in the previous section actually has a redundancy.
By properly dealing with the redundancy one can slightly reduce the cost.
A key point is that one does not have to reconstruct the network formed by four-leg tensors
in every coarse-graining step.
In the following we will explain such an algorithm, and we call it PTTRG2.

\begin{figure}[t]
\centering
 \includegraphics[width=80mm]{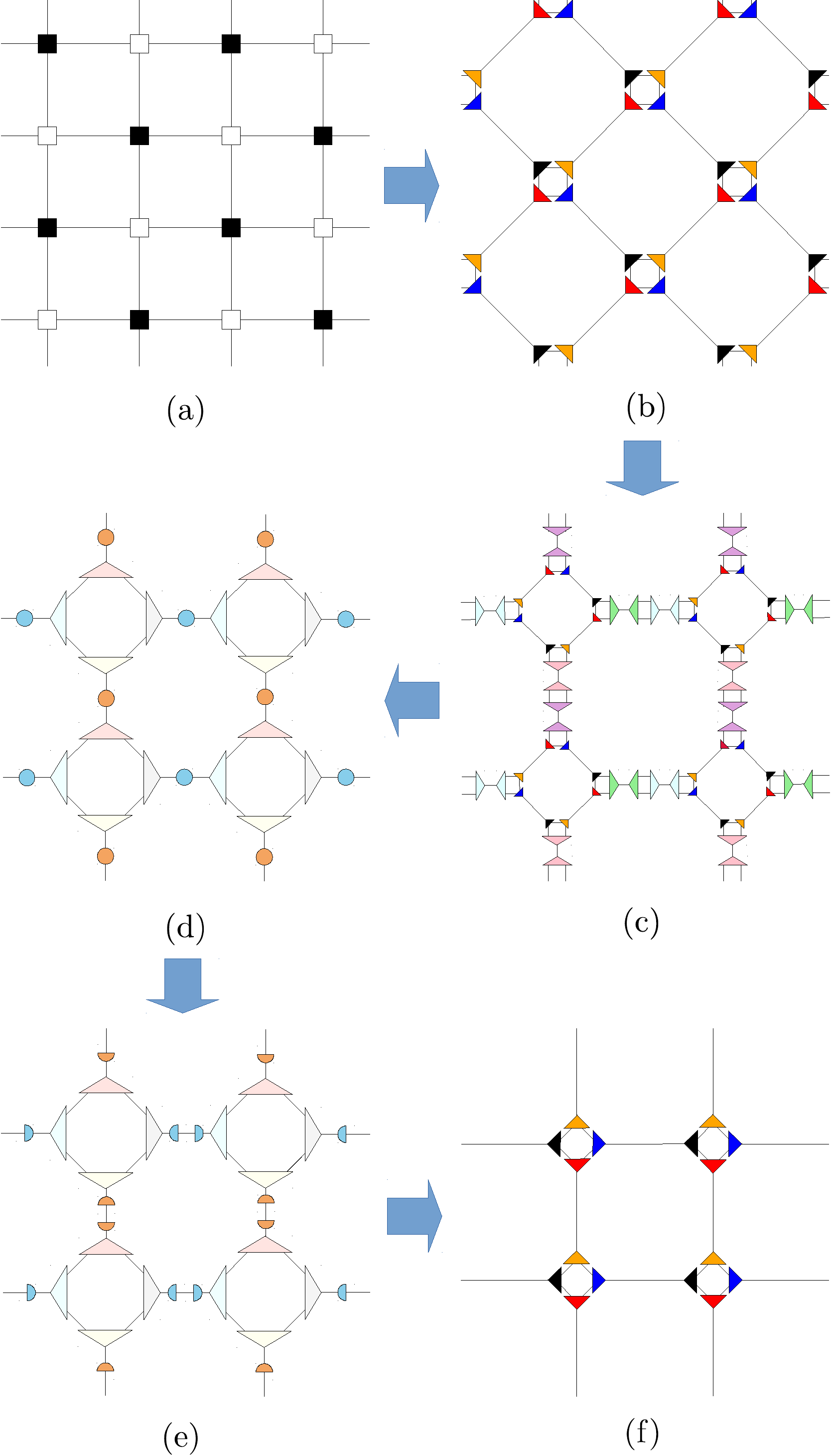}
 \caption{Flow of PTTRG2.
 From (a) to (b) decomposing a four-leg tensor to two three-leg tensors.
 From (b) to (c) insertion of two kinds of projectors for each direction.
 From (c) to (d) local contractions are done. A contraction of two isometries gives a matrix.
 Contracting two three-leg tensors with an isometry gives a three-leg tensor.
 From (d) to (e) decomposing the matrices.
 From (e) to (f) contraction between the three-leg tensor
 and the decomposed matrix gives a new three-leg tensor.
 The network formed by 3-leg tensors $S^{(1,2,3,4)}$ is constructed.
 Then we go back to (b) and repeat the coarse graining.
 }
 \label{fig:pttrg2}
\end{figure}

The flow of the PTTRG2 algorithm is shown in Fig.~\ref{fig:pttrg2}.
Let us begin with Fig.~\ref{fig:pttrg}(a) for PTTRG,
which is again shown in Fig.~\ref{fig:pttrg2}(a).
In the process from Fig.~\ref{fig:pttrg2}(a) to Fig.~\ref{fig:pttrg2}(b) 
where the four-leg tensor is decomposed into two three-leg tensors\footnote{This process is done once, that is,
only in the first iteration of the coarse graining.},
one may use the same procedure as the PTTRG decomposition, that is, from Fig.~\ref{fig:pttrg}(a) to Fig.~\ref{fig:pttrg}(e).
The next processes from Fig.~\ref{fig:pttrg2}(b) to Fig.~\ref{fig:pttrg2}(d) are also done
in the same way as those from Fig.~\ref{fig:pttrg}(e) to Fig.~\ref{fig:pttrg}(g) for PTTRG.
A key step is from Fig.~\ref{fig:pttrg2}(d) to Fig.~\ref{fig:pttrg2}(e), where
instead of making the four-leg tensor [PTTRG, Fig.~\ref{fig:pttrg}(g) to Fig.~\ref{fig:pttrg}(h)],
the matrix (circle) in Fig.~\ref{fig:pttrg2}(d)
is decomposed\footnote{We use the SVD for this decomposition.}
into two matrices [Fig.~\ref{fig:pttrg2}(e)].
Actually, the redundant part in PTTRG is that forming a new four-leg tensor
from the two three-leg tensors and the matrix in Fig.~\ref{fig:pttrg}(g) to Fig.~\ref{fig:pttrg}(h).
Since the new tensor will be decomposed into two three-leg tensors in
the next coarse-graining step anyway,
one does not have to make the four-leg tensor.
Then in Fig.~\ref{fig:pttrg2}(f) one makes three-leg tensors $S^{(1,2,3,4)}$.
By returning to Fig.~\ref{fig:pttrg2}(b) with the rotation, one can restart the coarse graining 
of the network formed by the three-leg tensors.
Note that during the process in Fig.~\ref{fig:pttrg2}(b) to Fig.~\ref{fig:pttrg2}(f),
one needs only the four projectors for the PTTRG contraction,
thus the computational cost can be reduced compared with PTTRG.

Here let us mention an equivalence of PTTRG and PTTRG2.
As explained at the end of Sec.~\ref{subsec:PTTRG},
PTTRG essentially performs a single information reduction from $\chi^2$ to $\chi$
in a coarse graining-step.
For PTTRG2, such a reduction is done only once as well 
since there is no process for a decomposition of a four-leg tensor.
In this way, PTTRG and PTTRG2 have a common feature
that performs the single information reduction;
thus, one sees that PTTRG and PTTRG2 are essentially equivalent,
and
it is natural to expect that the result of PTTRG agrees with that of PTTRG2
for sufficiently large $n_\itr$.

\subsection{Cost-reduced PTTRG \label{subsec:PTTRG3}}
The projective truncation technique is a very universal approach
and can be applied to any network.
Here we propose an algorithm (PTTRG3) which aggressively reduces the cost;
that is, the pre factor of $\chi^5$ is made much smaller.

\begin{figure}[t]
\centering
 \includegraphics[width=80mm]{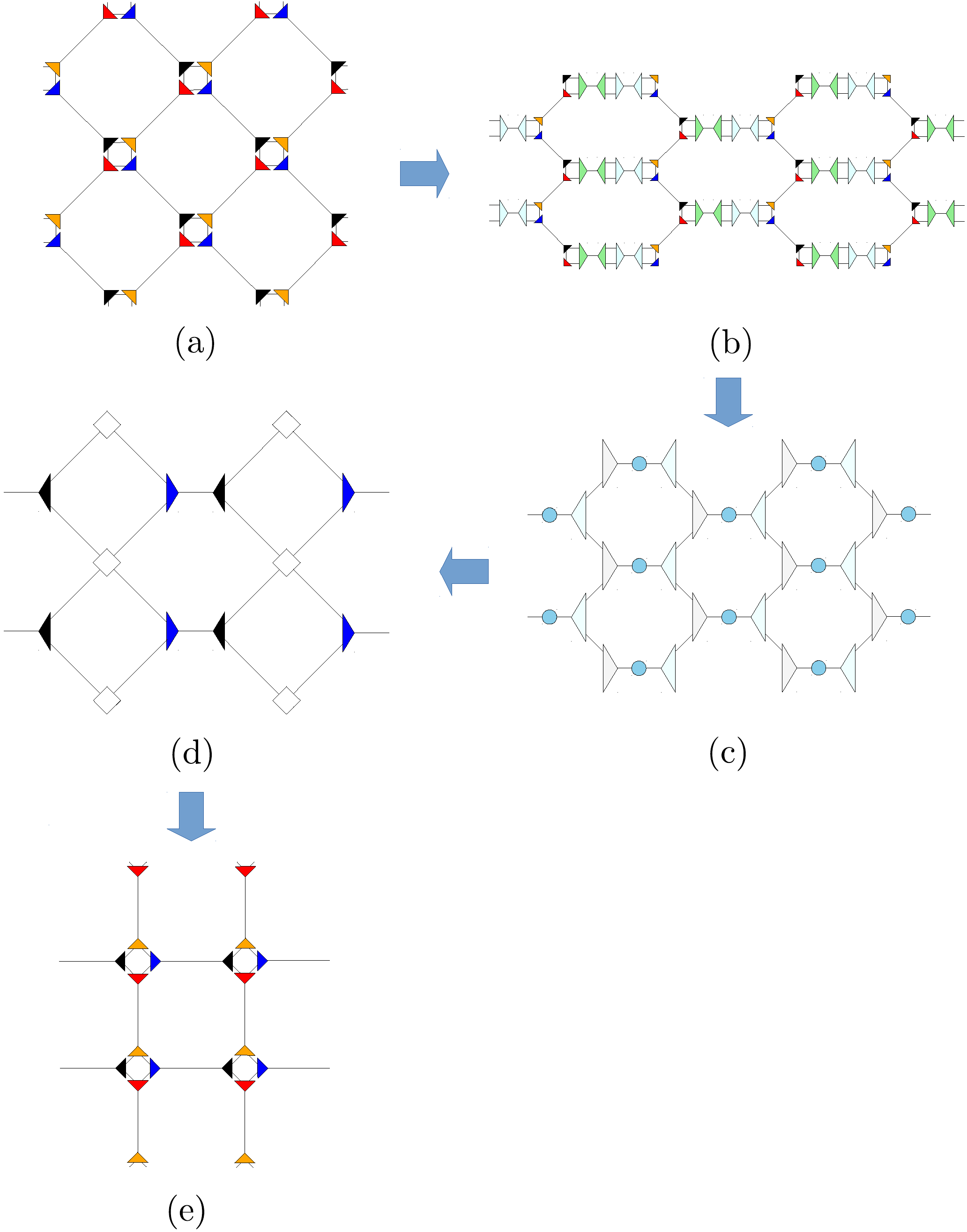}
 \caption{PTTRG3 algorithm.
   (a) Network consisting of three-leg tensors $S^{(1,2,3,4)}$.
   From (a) to (b) two projectors, $aa^\dag$ and $bb^\dag$, are inserted for the horizontal direction.
   From (b) to (c) contracting an isometry and two three-leg tensors gives a three-leg tensor.
   Contraction between $a$ and $b^\dag$ gives a matrix (blue circle).
   From (c) to (d) in one group, contracting the matrix and two three-leg tensors gives a four-leg tensor.
   In the other group, after decomposing the matrix, a contraction between
   the decomposed matrix and the three-leg tensor gives a new three-leg tensor, $S^{(1)}$ or $S^{(3)}$.
   From (d) to (e) for the former group, 
   the four-leg tensor is decomposed into two three-leg tensors, $S^{(2)}$ and $S^{(4)}$.
   Then we go back to (a).}
 \label{fig:pttrg3}
\end{figure}

The flow of PTTRG3 is shown in Fig.~\ref{fig:pttrg3}.
Let us begin with Fig.~\ref{fig:pttrg3}(a),
 whose network is formed by the three-leg tensors $S^{(1,2,3,4)}$.
From Fig.~\ref{fig:pttrg3}(a) to Fig.~\ref{fig:pttrg3}(b), the two projectors are inserted
only for the horizontal direction.
From Fig.~\ref{fig:pttrg3}(b) to Fig.~\ref{fig:pttrg3}(c), local contractions are done
and then a matrix (circle) and two three-leg tensors appear.
The process from Fig.~\ref{fig:pttrg3}(c) to Fig.~\ref{fig:pttrg3}(d) is a little bit tricky.
We consider two groups in the network: in one group the matrix and two three-leg tensors are contracted
to make a four-leg tensor, and in the other group the matrix is decomposed and then
the contraction between the decomposed matrix and a three-leg tensor gives $S^{(1)}$ or $S^{(3)}$.
From Fig.~\ref{fig:pttrg3}(d) to Fig.~\ref{fig:pttrg3}(e), the four-leg tensor is decomposed into two three-leg tensors,
$S^{(2)}$ and $S^{(4)}$.
Note that this decomposition is done only for the vertical direction.
In this way, one can reduce the number of projectors for both the contraction and decomposition parts
by half compared with PTTRG; thus,
the total cost of PTTRG3 is expected to be nearly half that of PTTRG.

\begin{figure}[t]
\centering
\includegraphics[width=80mm]{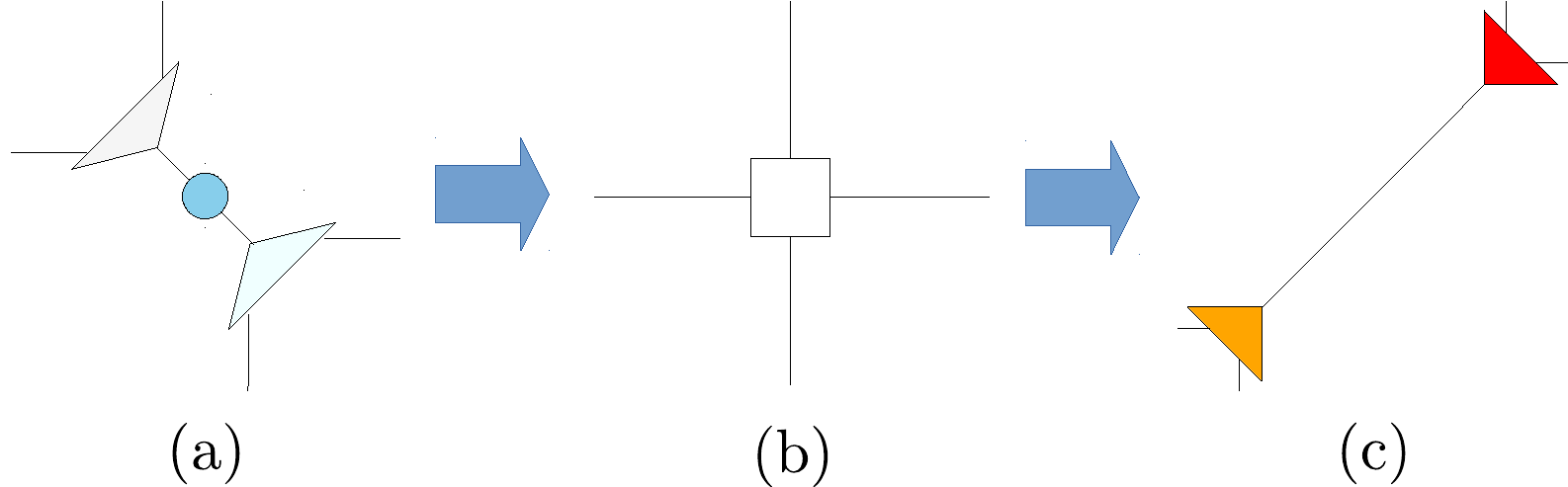}
 \caption{The direction of the contraction is different from that of the decomposition.}
 \label{fig:pttrg_lr}
\end{figure}

There is an important point for the PTTRG3 algorithm.
In the process of Fig.~\ref{fig:pttrg3}(c) to Fig.~\ref{fig:pttrg3}(d) to Fig.~\ref{fig:pttrg3}(e),
for the second group, the direction of the contraction
is different from that of the decomposition,
as shown in Fig.~\ref{fig:pttrg_lr}.
This indicates that, 
in addition to the information reduction from $\chi^2$ to $\chi$ [from Fig.~\ref{fig:pttrg3}(a) to Fig.~\ref{fig:pttrg3}(c)],
another reduction of information
may happen in the decomposition process from Fig.~\ref{fig:pttrg_lr}(b) to Fig.~\ref{fig:pttrg_lr}(c),
while there is no such information loss in the process from Fig.~\ref{fig:pttrg_ll}(b) to Fig.~\ref{fig:pttrg_ll}(c) in PTTRG.
Therefore, it is not guaranteed that PTTRG3 converges to PTTRG (or, equivalently, to TRG)
even for a large number of iterations.

\subsection{Comparison of computational time}
Here let us roughly compare the expected computational time for the three algorithms.
In the following, we assume that the number of iterations
$n_{\rm itr}$ and the bond dimension $\chi$ are sufficiently large.
In such a circumstance,
the total time of the PTTRG algorithm $T_{\rm PTTRG}$ is dominated
by the determination of isometries for the PTTRG decomposition ($T_{\rm D}$) and
the contraction parts ($T_{\rm C}$),
\begin{align}
T_{\rm PTTRG}
\sim
T_{\rm D}
+
T_{\rm C}.
\end{align}
By analyzing the number of operations for the two parts, we find that
\begin{align}
T_{\rm C}
\sim
2
T_{\rm D}.
\end{align}
For the PTTRG2 algorithm, as explained in Sec.~\ref{subsec:PTTRG2},
the projectors in the PTTRG decomposition part are not required; 
thus, the total time is estimated as
\begin{align}
T_{\rm PTTRG2}
\sim
T_{\rm C}.
\end{align}
As for the PTTRG3 algorithm, 
since one needs projectors for one direction in both the decomposition and contraction parts
as explained in Sec.~\ref{subsec:PTTRG3},
the total time is expected to be half that of PTTRG,
\begin{align}
T_{\rm PTTRG3}
\sim
T_{\rm PTTRG}/2.
\end{align}
In summary, one finds
\begin{align}
T_{\rm PTTRG}
:
T_{\rm PTTRG2}
:
T_{\rm PTTRG3}
=
1
:
\frac{2}{3}
:
\frac{1}{2}.
\end{align}

\section{Numerical results \label{sec:numerical_results}}
In this section, we present the result of the two numerical experiments:
the free energy of the Ising model on the square lattice and the elapsed time
of our algorithms and TRG.

\begin{figure}[t]
  \centering
  \includegraphics[width=80mm]{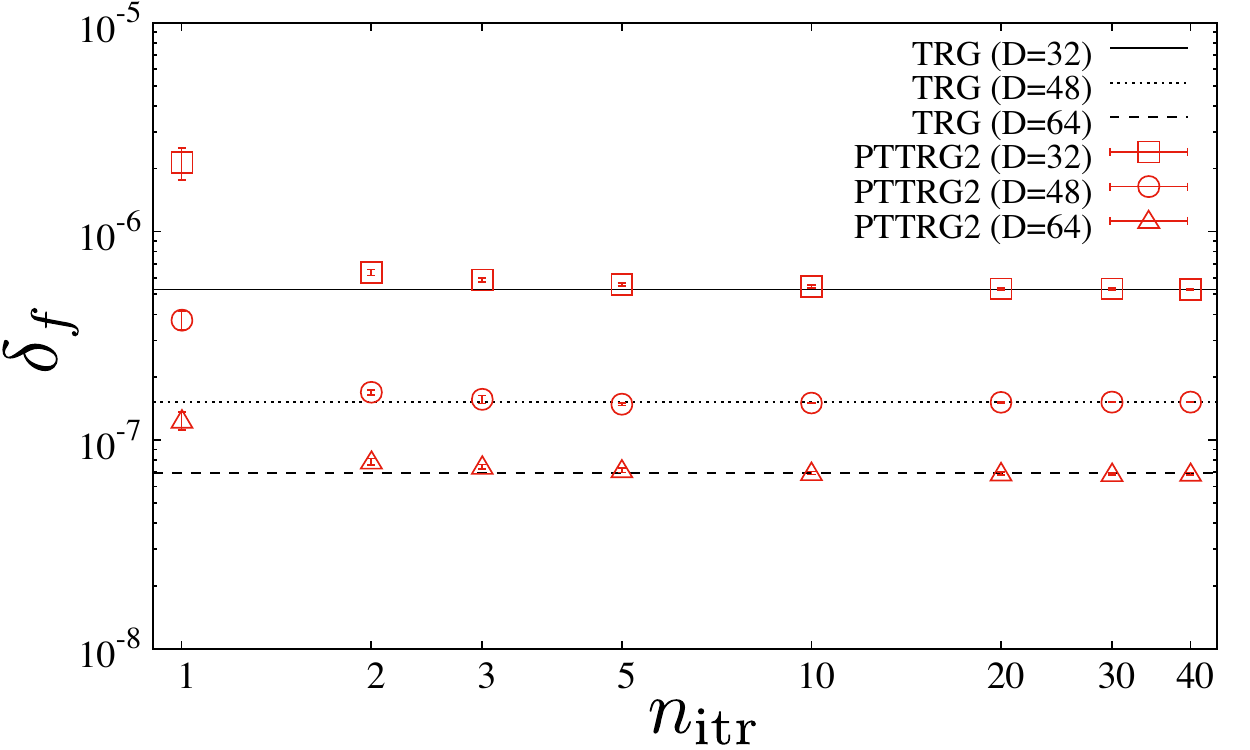}
  \caption{The relative error of the free energy as a function of the iteration number $n_\itr$
  at $T_\cc$ with $2^{30}$ spins for $\chi=32$, $48$ and $64$.
  PTTRG2 reaches to the precision of TRG with $n_\itr\gtrsim 5$.
  }
  \label{fig:pttrg2_T_Tc}
\end{figure}

To see the accuracy of the physical quantity for the algorithms, we compare
the relative error of the free energy defined by
\begin{align}
  \delta_f = \left|\frac{f-f_{\mathrm{exact}}}{f_{\mathrm{exact}}} \right|.
\end{align}
Figure \ref{fig:pttrg2_T_Tc} plots $\delta_f$ as a function of
the iteration number $n_\itr$ for the determination of the isometries 
at the critical temperature $T_\cc$ on a $2^{15}\times2^{15}$ lattice, i.e.,
at 30 coarse-graining steps.
Note that the values of $n_\itr$ for the determination of all isometries are taken to be the same,
and they are fixed during the coarse-graining steps.
We prepare 15 initial isometries, and the
error bars shown in Fig.~\ref{fig:pttrg2_T_Tc} are the standard deviation estimated by using the 15 samples.
We find that PTTRG2 achieves the accuracy of TRG 
after a few iteration steps
for all the bond dimensions we investigated, $\chi=32$, $48$, and $64$.
Furthermore, we observe that the value of $n_\itr$, where $\delta_f$ saturates,
is independent of the bond dimensions.
Note that the error bars rapidly get smaller with increasing $n_\itr$
since the initial value dependence of the isometry
is reduced for larger iteration steps.

\begin{figure}[t]
  \centering
  \includegraphics[width=80mm]{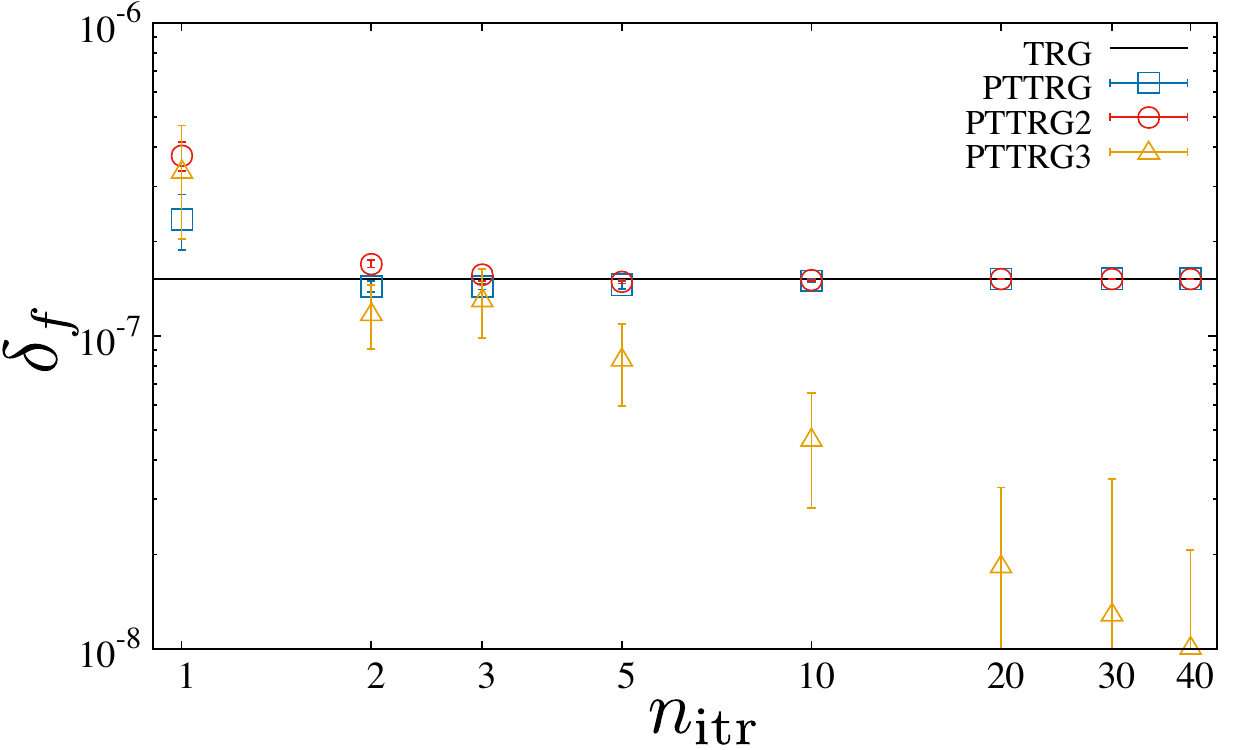}
  \caption{The relative error of the free energy for new algorithms and TRG
    at $T_\cc$ with $2^{30}$ spins for $\chi=48$.}
  \label{fig:D_48_T_Tc}
  \includegraphics[width=80mm]{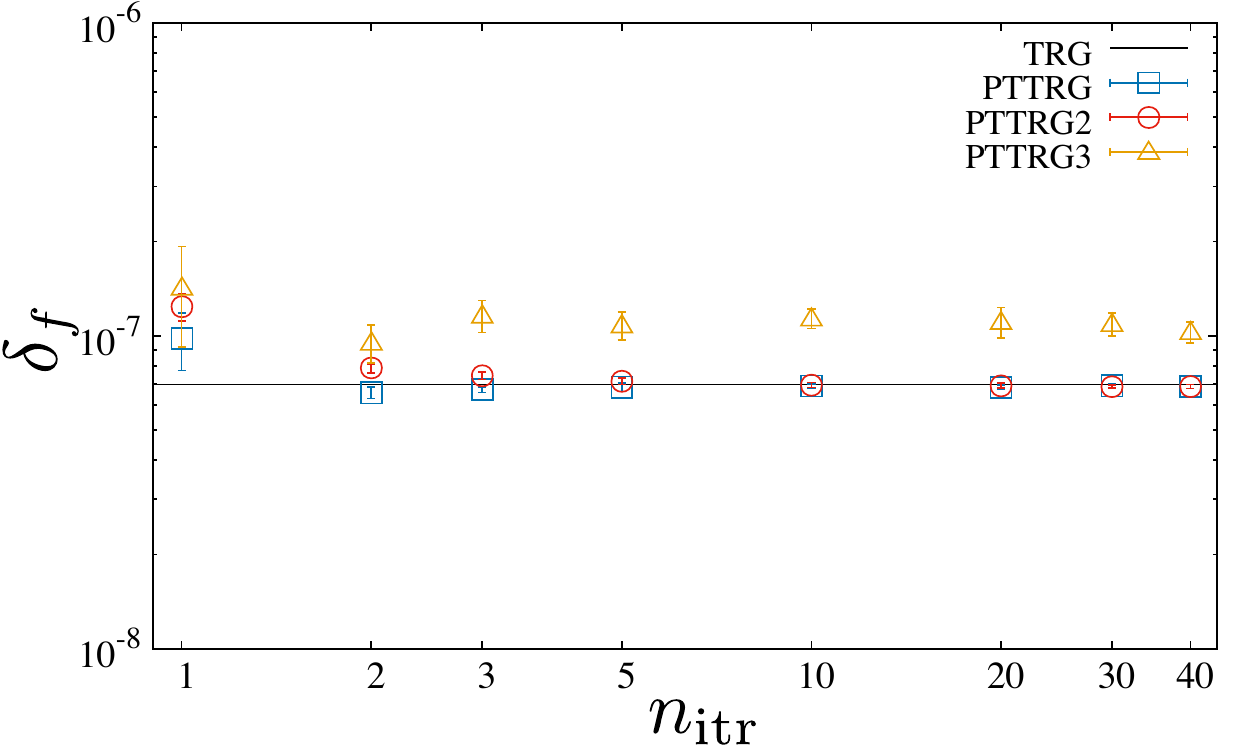}
  \caption{The relative error of the free energy for new algorithms and TRG
    at $T_\cc$ with $2^{30}$ spins for $\chi=64$.}
  \label{fig:D_64_T_Tc}
\end{figure}

Figure \ref{fig:D_48_T_Tc} shows $\delta_f$
for our algorithms (PTTRG, PTTRG2, and PTTRG3) and TRG at $T_\cc$ with $\chi=48$.
We find that both the PTTRG and PTTRG2 results smoothly
converge to that of TRG with increasing $n_\itr$,
and this tendency is also seen for $\chi=64$, 
as shown in Fig.~\ref{fig:D_64_T_Tc}.
This behavior is expected, 
as explained at the end of Secs.~\ref{subsec:PTTRG} and \ref{subsec:PTTRG2}.
The PTTRG3 result, however, exhibits behavior different from that of PTTRG and PTTRG2.
For $\chi=48$, it is somehow more accurate
than TRG (Fig.~\ref{fig:D_48_T_Tc}).
On the other hand, with the larger bond dimension $\chi=64$,
such a tendency is not observed anymore,
and the accuracy of the free energy gets worse,
as shown in Fig.~\ref{fig:D_64_T_Tc}. 
Furthermore, the error bars remain visible even for relatively larger $n_\itr$;
that is, the effect of the initial isometry persists.
	A possible reason for why the results of PTTRG3 are unstable is that
	the algorithm experiences two information reduction steps
	per coarse graining, as explained in Sec.~\ref{subsec:PTTRG3},
	although PTTRG and PTTRG2 have a single reduction step. 
	This additional reduction may cause the iterative method
	to fail to attain a minimum, and the resulting isometry is not the best one.
	Therefore, we conclude that PTTRG3 is not useful, although the cost is reduced aggressively.

\begin{figure}[t]
  \centering
  \includegraphics[width=80mm]{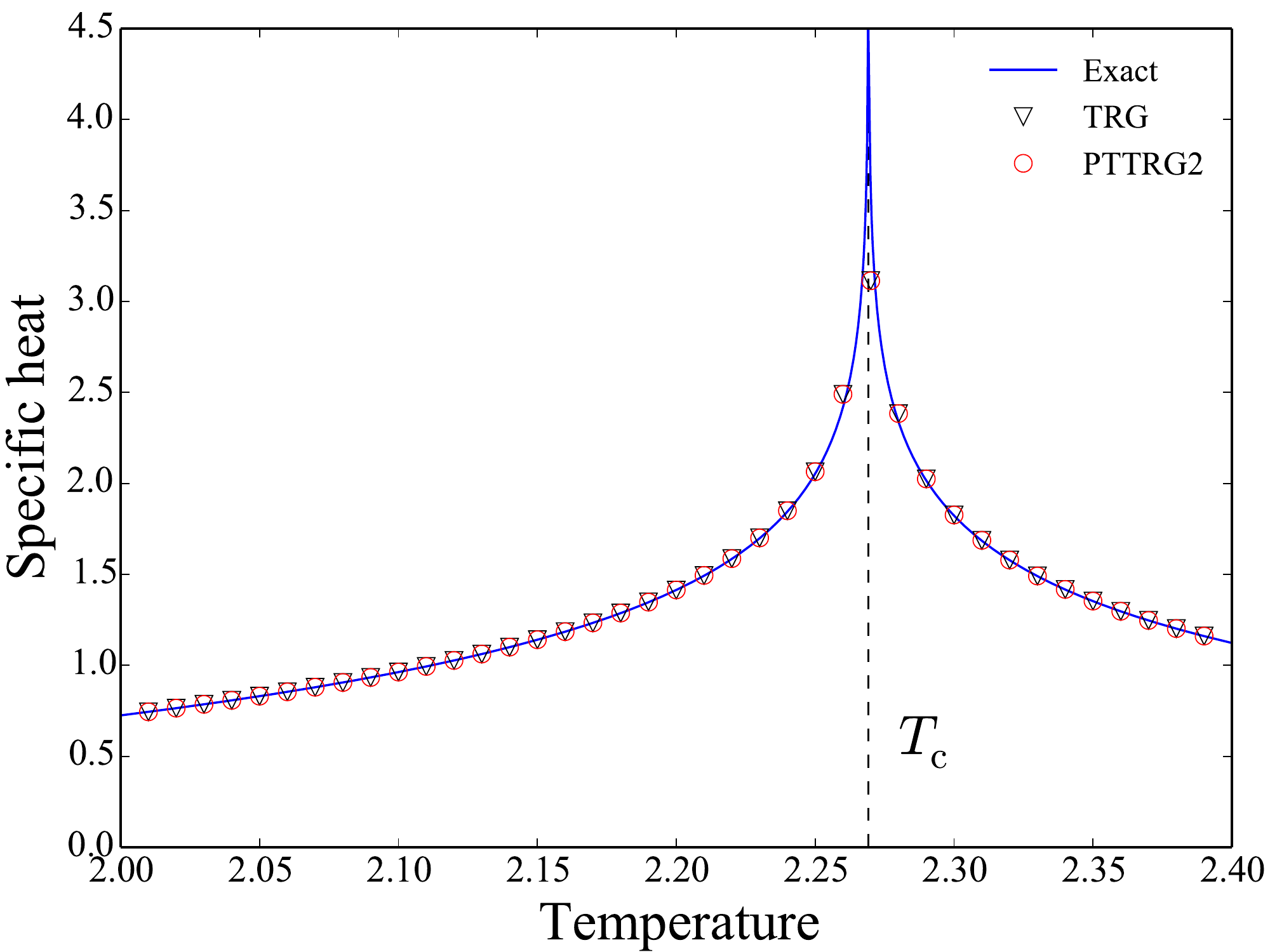}
  \caption{Specific heat around the critical temperature $T_\cc$ obtained by TRG and $n_\itr=10$ PTTRG2
    with $2^{30}$ spins for $\chi=32$.}
  \label{fig:specific_heat}
\end{figure}

As seen in Figs. \ref{fig:pttrg2_T_Tc}-\ref{fig:D_64_T_Tc}, the free energy of PTTRG and PTTRG2 reaches the precision
of the TRG; thus, thermodynamic functions obtained by the numerical derivatives
are also expected to maintain the same order of accuracy.
To confirm this, we study the specific heat obtained by numerical derivatives of the free energy.
Figure \ref{fig:specific_heat} shows the specific heat 
around the critical temperature $T_\cc$ using TRG and PTTRG2.
We use the average values of the free energy for PTTRG2 with $n_\itr=10$.
As a result, the singular behavior around the critical point is clearly seen for both cases and
we see that the accuracy
of our algorithm is also comparable to that of TRG.

\begin{figure}[t]
\centering
 \includegraphics[width=80mm]{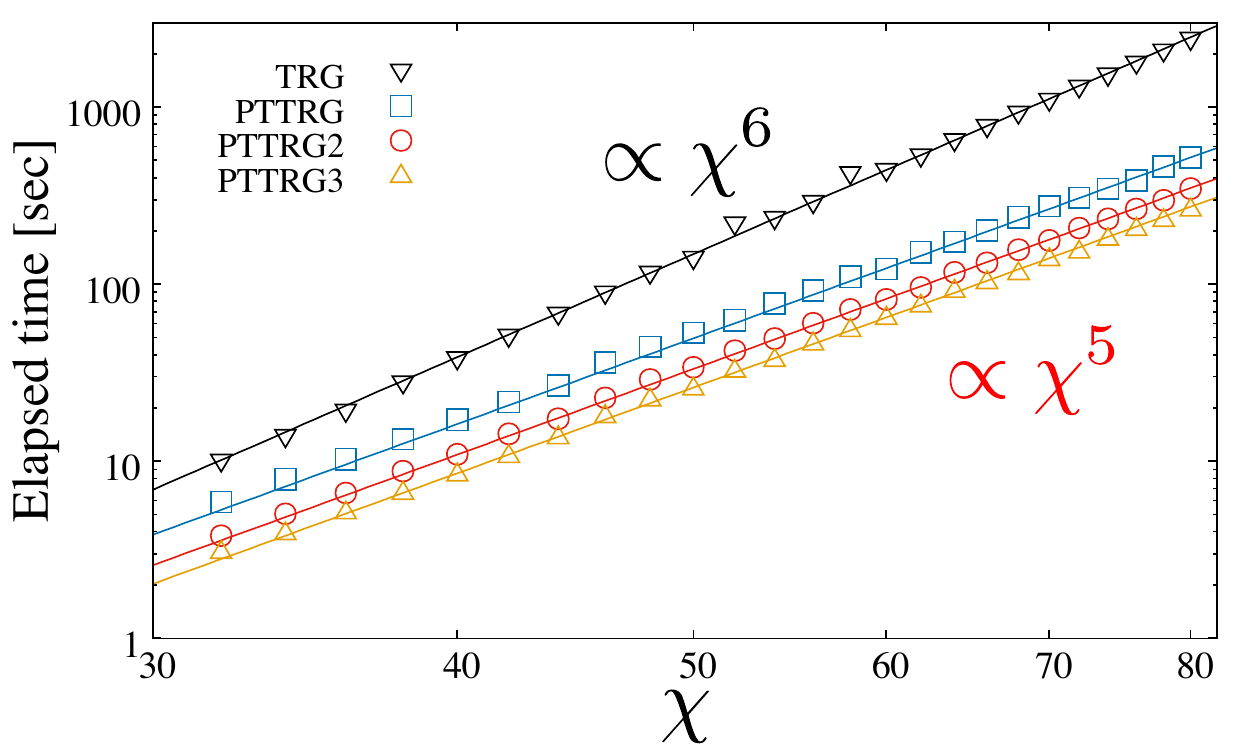}
 \caption{
       Elapsed time of the algorithms.
       The black line (TRG) is the fitting line $\propto \chi^6$.
       For the new algorithms ($n_\itr=10$), a fitting form is proportional to $\chi^5$.
 }
 \label{fig:time}
\end{figure}

Finally, let us see the elapsed time of these algorithms.
Figure~\ref{fig:time} shows the elapsed time per coarse graining at $T_\cc$
with $n_\itr=10$ for our algorithms.
From the Fig.~\ref{fig:time} we confirm that the cost of PTTRG scales with
$O(\chi^5)$ and that of TRG is $O(\chi^6)$.
The iteration number $n_\itr$ should be smaller than $\chi$;
otherwise, it does not become advantageous concerning the cost.
As seen in the relative error of the free energy, however,
PTTRG and PTTRG2 achieve the same accuracy as TRG with a low $n_\itr$;
thus, we have gained concerning the performance.

\section{Conclusions \label{sec:conclusion}}
We have explained three kinds of algorithms whose
concept is to reduce the computational cost by the
projective truncation technique
from $O(\chi^6)$ of TRG to $O(n_\itr\chi^5)$.
We also performed their numerical tests and,
indeed, confirmed the scaling by measuring the elapsed time.
We found that the accuracy of the free energy for the two-dimensional (2D) Ising model with our algorithms
is comparable to that of TRG with a few iteration steps.
Therefore we conclude that the new algorithms (PTTRG and PTTRG2) indeed have gained
compared with TRG.
PTTRGs can also treat 
the tensor network of 
the 2D classical spin model
which corresponds to 
a one-dimensional quantum spin model with Suzuki--Trotter decomposition
\cite{Suzuki--Trotter_1,Suzuki--Trotter_2},
and the cost is reduced to $O(\chi^5)$ as well.

Recently, Morita {\it et al.}~\cite{RSVDTRG} presented an $O(\chi^5)$ TRG algorithm
which applies the randomized SVD to the TRG decomposition part
in order to reduce the cost of the SVD to $O(\chi^5)$.
The cost of the TRG contraction part is also cleverly reduced to $O(\chi^5)$ without forming
the four-leg tensor, and its strategy is similar to that of our PTTRG2.
In contrast, our PTTRG uses isometries to reduce the cost
for both the decomposition and contraction parts independently\footnote{
Actually, we see the redundancy when the projective truncation technique
is applied to TRG.
We, however, think that this is a feature for TRG and
in general it is not expected.
}.
Therefore, we believe that the
projective truncation technique is more versatile.

Finally, we comment on the future perspective.
The projective truncation technique can be applied to any network to reduce the cost.
Therefore, for instance, one may reduce the cost of the higher-order TRG contraction part \cite{hotrg}
in higher-dimensional systems.
One should keep in mind, however, that the local approximation could be bad,
and it may affect the accuracy of physical quantities.
This issue depends on the target network itself,
and at the moment it seems hard to know the effectiveness of the method in advance.

\begin{acknowledgments}
The authors would like to thank S.~Morita for offering a PYTHON code of TRG.
The computation in this work was carried out on computers at RIKEN R-CCS.
This work is supported in part by
JSPS KAKENHI Grant No. JP17K05411
and
MEXT as ``Exploratory Challenge on Post-K computer'' (Frontiers of Basic Science: Challenging the Limits).
\end{acknowledgments}

\appendix
\section{Determination of Isometry \label{sec:det_iso}}
In this appendix we summarize how to determine isometries \cite{VidalEvenbly,isometry} in PTTRG.
For a given network ${\cal N}$,
we want to find an optimal isometry $w$ which satisfies
\begin{align}
  \min_{w} \delta &= \min_{w} \frac{||{\cal N} - {\cal N}ww^\dag|| \label{eq:delta}}{||{\cal N}||}.
\end{align}
The cost function $\delta$ can be deformed as
\begin{align}
  \delta^2 &= 1 
  - \frac{||{\cal N}w||^2}{||{\cal N}||^2}.
\end{align}
To minimize $\delta$, we have to solve the problem
\begin{align}
  \max_{w} &||{\cal N}w||^2
  = \max_{w} \mathrm{Tr}\Gamma_{w} w\label{eq:w1w2},
\end{align}
where $\Gamma_{w}$ is called an environment for $w$.
Fixing $w^\dag$ in $\Gamma_w$, we can treat the right-hand side of Eq. (\ref{eq:w1w2})
as a linear problem with respect to $w$.
The environment $\Gamma_w$ has the same partition of indices as $w$.
When it is considered as a matrix, its SVD is given by
\begin{align}
  \Gamma_{w} = usv^\dag,
\end{align}
where the singular values are ordered $s_1\ge s_2\ge ...\ge s_\chi\ge0$ and
$v$ and $u$ are $\chi^2\times\chi^2$ and $\chi\times\chi$
unitary matrices, respectively, if $w$ is a $\chi^2\times\chi$ matrix.
In fact, the optimal solution for the problem on the right-hand side of Eq. (\ref{eq:w1w2}) is given by
\begin{align}
  w = v'u^\dag,
\end{align}
where $v'$ is a $\chi^2 \times \chi$ matrix made of the first $\chi$ column vectors of $v$.
Next, $w^\dag$ is replaced by using the updated $w$, 
and usually, one repeats the steps
until $w$ is sufficiently converged.
We define $n_\itr$ as the number of iteration steps.
To start the iteration, an initial isometry is chosen randomly\footnote{
As an initial isometry, one can use an isometry determined
in a previous renormalization group step and it may significantly reduce
the computational time.
}
under the constraint $w^\dag w=\mathbb{I}$.

\begin{figure}
\centering
 \includegraphics[width=80mm]{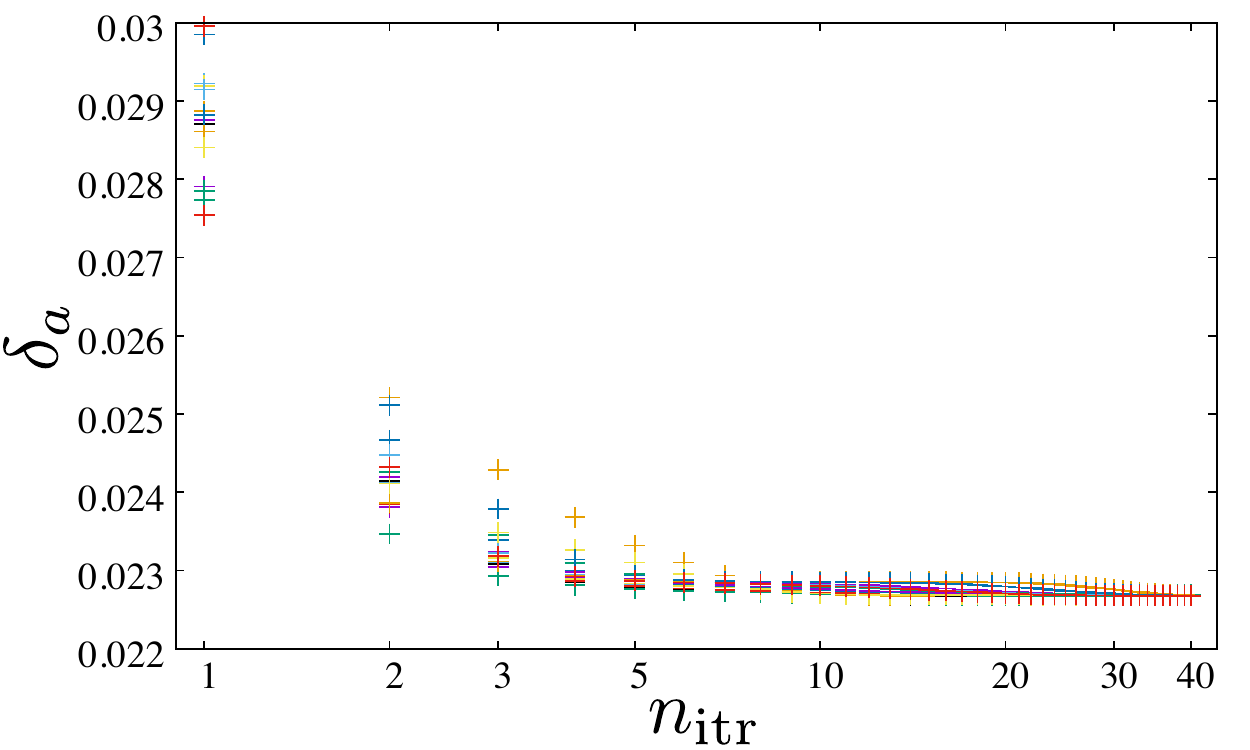}
 \caption{The cost function $\delta_a$ in Eq. (\ref{eq:delta_a})
 as a function of iteration number $n_{\rm itr}$ on the $2^5$ lattice at $T_\cc$.}
 \label{fig:notdec_w1}
\end{figure}

Figure \ref{fig:notdec_w1} shows $\delta_a$ in Eq. (\ref{eq:delta_a}) of the PTTRG contraction part
at the fifth coarse-graining step\footnote{
A non trivial truncation by the projectors  starts to appear from
the fifth coarse-graining step
in the case of $\chi\ge32$ for the Ising model on the square lattice.
} ($2^5$ lattice) for the Ising model on the square lattice at the critical temperature $T_\cc$.
In Fig.~\ref{fig:notdec_w1}, points associated with 15 initial isometries are superimposed.
After $n_{\rm itr}\sim10$, the value of $\delta_a$ saturates and
the dependence on the initial value disappears.

\newpage



\end{document}